# A framework for modeling aerosol-cloud-lightning interactions: Validation of charge structure and aerosol effects

**Weishan Wang[1], Guoxing Chen[1,2], Yijun Zhang[1,2], Jen-Ping Chen[3], Dong Zheng[4,5], Liangtao Xu[4,5]**

[1]Department of Atmospheric and Oceanic Sciences, Institute of Atmospheric Sciences, and CMA-FDU Joint Laboratory of Marine Meteorology, Fudan University, Shanghai, China

[2]Shanghai Frontiers Science Center of Atmosphere-Ocean Interaction, Fudan University, Shanghai, China

[3]Department of Atmospheric Sciences, National Taiwan University, Taipei, Taiwan

[4]State Key Laboratory of Severe Weather Meteorological Science and Technology

[5] CMA Key Laboratory of Lightning, Chinese Academy of Meteorological Sciences, Beijing, China

Corresponding author: Guoxing Chen (chenguoxing@fudan.edu.cn)

**Key Points:**

- An aerosol-cloud microphysics-lightning coupled framework was developed in the WRF model to study their interactions in thunderstorms.
- The framework successfully reproduces the observed spatiotemporal evolution of the tripolar positive-negative-positive charge structure.
- Aerosols enhance lightning by increasing cloud ice concentration, which boosts graupel-ice collisions and intensifies charge separation.

**Abstract**

This study develops a novel framework within the Weather Research and Forecast Model for modeling aerosol-cloud-lightning interactions. The framework explicitly represents aerosol-cloud interactions by prescribing aerosols with two configurations: an idealized setup, where both cloud condensation nuclei (CCN) and ice nucleating particles (IN) are assumed to have a single chemical composition and spatially uniform distributions; and a quasi-realistic configuration, with multi-species aerosols assigned spatially varying distributions, where hygroscopic components act as CCN, dust particles act as IN, and all aerosol species influence radiative transfer. Cloud microphysics is coupled with detailed charge separation and discharge processes to enable the lightning simulation. The framework is evaluated using two thunderstorms in Guangdong, China. For an isolated storm, the model successfully reproduces the observed tripolar charge structure (positive–negative–positive), demonstrating its capability in simulating cloud electrification. For a frontal storm, it captures well the observed precipitation and lightning, and shows that increasing CCN suppresses the rainfall while enhancing the lightning. Higher CCN concentrations produce more numerous but smaller cloud droplets, which suppresses the coalescence into rain droplets, allows a greater number of droplets to loft into the upper troposphere, and forms more but smaller cloud ice particles. This boosts graupel–ice collisions, intensifies non-inductive charging, strengthens the upper positive charge and the vertical electric-field gradient, ultimately increasing the lightning frequency. In contrast, no significant aerosol-induced invigoration of updrafts is observed. These results highlight the dominant role of aerosol microphysical effects over dynamical invigoration in modulating thunderstorm electrification and lightning activity.

**Plain Language Summary**

Lightning is a destructive weather phenomenon responsible for thousands of fatalities annually. It primarily occurs in deep convective systems and thus serves as an indicator of convective intensity. Previous studies have shown that aerosols can modulate lightning activity by altering cloud microphysical processes; however, the underlying mechanisms remain debated due to limitations in existing models in representing aerosol–cloud–lightning interactions. To address this gap, this study develops a modeling framework within the Weather Research and Forecasting (WRF) Model to simulate these coupled processes and evaluates its performance through comparisons with observations and sensitivity experiments. Results demonstrate that the model accurately reproduces the vertical structure of the electric field, surface precipitation, and lightning activity, closely matching observations. It also captures their responses to varying aerosol concentrations. The enhancement of lightning is primarily attributed to increased ice–graupel collisions under higher aerosol loading, which amplifies non-inductive charge separation and strengthens the vertical electric-field gradient. In contrast, the effect of aerosol-induced convective invigoration is not evident.

**1 Introduction**

Lightning, a hazardous meteorological phenomenon, plays an important role in the Earth's climate system. It alters atmospheric chemical composition through producing nitrogen oxides (Grewe, 2007; Schumann & Huntrieser, 2007; Ott et al., 2010) and affects terrestrial ecosystems by igniting wildfires (Krause et al., 2014; Pérez-Invernón et al., 2023). Lightning activity is closely tied to deep convective processes, making it a reliable indicator for the convection intensity (Rutledge et al., 1992; Petersen et al., 1999; Ávila et al., 2010). The global lightning activity has been observed to increase in the recent decade (Saha et al., 2017; Qie et al., 2021), yet its future changes within the context of global warming remain debated (Romps et al., 2014; Finney et al., 2018). Therefore, studying lightning is of considerable scientific importance for improving our understanding of atmospheric processes, enhancing weather forecasting capabilities, and assessing environmental risks.

Factors controlling lightning activity can roughly be grouped into two categories (Williams 1989; Carey et al., 2003; Wang et al., 2018). The first category is thermodynamic conditions that determine the macroscale organization and lifecycle characteristics of convective systems (Stolz et al., 2015). For example, large convective available potential energy (CAPE) and weak vertical wind shear show strong positive correlations with the lightning activity, while warm cloud depth and low environmental relative humidity are negatively correlated with the lightning activity (Stolz et al., 2017). The second category is cloud microphysical processes (Yang et al., 2024), which modulate lightning activity through electrification mechanisms. The primary charge separation in thunderstorms arises from non-inductive collisions between graupel and cloud ice in the mixed-phase region (-5°C to -20°C). Enhanced supercooled liquid water content improves charge separation efficiency through accelerated graupel riming growth (Saunders & Peck, 1998) and secondary ice production via the Hallett-Mossop process (Hallett & Mossop, 1974). These combined mechanisms promote the development of charge structures, notably the 'bottom-heavy tripole' configuration (Bruning et al., 2007), ultimately giving rise to lightning discharges (Mansell et al., 2010).

Aerosols can act on the above two categories of factors to modulate lightning activity. First, aerosols can strengthen low-level inversions and atmospheric stability by absorbing and scattering the solar radiation, which suppresses the development of convective systems (Wang et al., 2018) and lightning activity. This has been consistently verified in thunderstorm activity across different climate regimes, such as the Arabian Peninsula (Dayeh et al., 2021) and the Pearl River Delta region (Zhang et al., 2025). Second, aerosols can interfere with cloud microphysical processes by serving as cloud condensation nuclei (CCN) and ice nuclei (IN), which affects concentrations of cloud droplets and ice-phase particles (Twomey, 1977; Albrecht, 1989; Korolev et al., 2017; Zhao et al., 2019), and subsequently influence the convective intensity and charge separation processes. Increasing CCN concentration may suppress warm rain processes and enhance the upward transport of supercooled water, forming more ice particles in the upper troposphere and releasing more latent heat to ultimately boost the convection (Rosenfeld et al., 2008; Fan et al., 2018) and lightning frequency (Tan et al., 2016). This mechanism explains lightning anomalies in various observational events, including ship emissions (Thornton et al., 2017; Wright et al., 2025) and volcanic eruptions (Yuan et al., 2011). Nevertheless, the mechanism of the aerosol invigoration effect is still debated in observations and simulations (Öktem et al., 2023; Varble et al., 2023), and there are also studies indicating that the increase of coarse aerosols may inhibit lightning as they accelerate warm rain at the expense of mixed-phase rain (e.g., Pan et al., 2022).

Numerical models have been serving as essential tools for understanding aerosol-cloud-lightning interactions. On the one hand, the foundational work began with Pringle et al. (1973) pioneering the incorporation of explicit electrification mechanisms into cloud models. Rawlins

(1982) achieved a three-dimensional cloud-lightning coupling model. By incorporating more observed mechanisms (e.g., Takahashi, 1978; Ziegler et al., 1991; MacGorman et al., 2001), Mansell et al. (2005) systematically compare multiple parameterizations of noninductive graupel-ice charging derived from laboratory studies. Their work identified the specific temperature and moisture regimes under which this mechanism dominates storm electrification, while elucidating the secondary role of inductive charging. The Fierro research team (Fierro & Reisner, 2011, 2013) further advanced lightning simulations in severe convective systems like hurricanes, revealing the connection between eyewall lightning and storm intensification. Subsequent studies have successfully simulated both the spatiotemporal lightning characteristics (Xu et al., 2014; Sun et al, 2021) and the three-dimensional evolution of charge structures (Popová et al., 2023). On the other hand, the representation of aerosol-cloud interactions in models has been continuously improved (Thompson et al., 2014; Cheng et al., 2010; Morrison et al., 2005). Notably, current models capture the mechanisms that increasing CCN concentration may change the cloud droplet spectral distribution, enhance latent heat release, promote updraft intensification, and ultimately affect the precipitation. However, whether these mechanisms lead to the convective invigoration depends heavily on the specific meteorological conditions. Furthermore, these models depict how elevated ice-nucleating particle (INP) concentrations modulate mixed-phase clouds: higher IN concentrations lead to increased cloud crystal number concentrations, consequently reducing supercooled liquid water within the cloud (Wang et al., 2024). This alters latent heat release and drives cloud dynamic adjustments.

However, several issues remain challenging for current numerical models. First, existing studies on aerosol-cloud interactions and cloud-lightning relationships are routinely conducted in isolation, creating a critical knowledge gap in understanding the aerosol-cloud microphysics-lightning chain. Second, the lightning models usually oversimplify the representation of aerosols (e.g., assuming fixed spatiotemporal concentrations and neglecting size-resolved distributions and composition-dependent activation processes). Third, most lightning models ignore aerosol-radiation interactions, thus overlooking their indirect effects on thunderstorm electrification and lightning activity.

Therefore, this study is motivated to construct a modeling framework that integrates lightning processes with an aerosol-aware cloud microphysical parameterization scheme within the Weather Research and Forecast (WRF) model. The framework simulates aerosol-radiation and aerosol-cloud interactions online using prescribed multi-component aerosol fields and explicitly represents the complete charging-to-discharging lightning processes, thus enabling quantitative assessment and mechanistic understanding of aerosol effects on convective electrification processes and lightning activities.

The remainder of this manuscript is organized as follows: Section 2 describes the coupled aerosol-cloud microphysics-lightning modeling framework. Section 3 introduces the data usage, the model configuration, and the experimental design. Section 4 validates the modeling framework performance for the charge structure, where an isolated thunderstorm over Guangdong, China, was simulated. Section 5 assesses aerosol CCN influences on lightning activity and decodes the underlying mechanism based on simulations of another thunderstorm case of Guangdong, followed by the summary and discussion presented in Section 6.

## 2 The framework for modeling Aerosol-cloud-lightning interactions

The framework for modeling aerosol–cloud–lightning interactions was implemented within the WRF model Version 3.7. As illustrated in Fig. 1, aerosols in the framework influence cloud microphysics by serving as CCN and IN while simultaneously directly modulating atmospheric radiation, which subsequently alters atmospheric instability and thereby affects cloud development and evolution. Both pathways can ultimately influence lightning activity. It is important to note that the coupling between cloud microphysics and lightning is unidirectional: microphysical processes can affect lightning processes (such as charge generation and discharge), whereas lightning processes do not feed back onto cloud microphysics.

Most submodules of the framework have been well documented in our previous studies (e.g., Cheng et al., 2007, 2010; Xu et al., 2014; Chen et al., 2018; Wang et al., 2024). In the following, we mainly provide a brief description of the model representation of aerosol–cloud and cloud–lightning interactions.

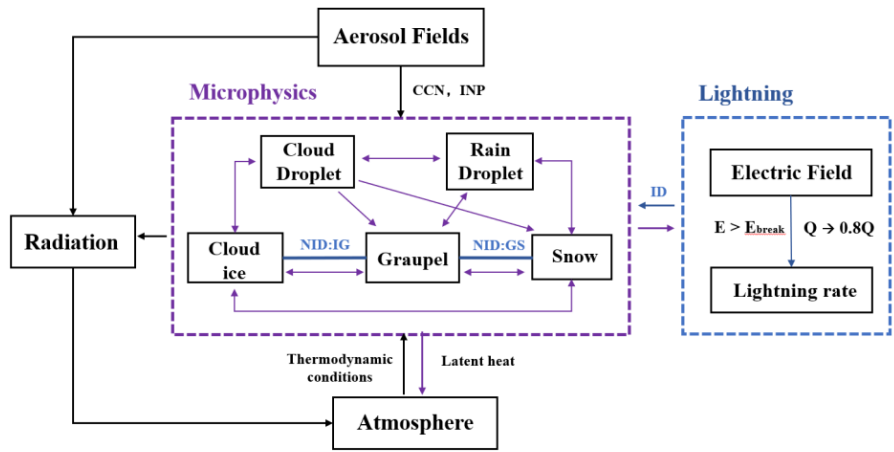

**Fig. 1.** Diagram of the framework for modeling aerosol-cloud microphysics-lightning interactions. ID and NID stand for inductive charging and non-inductive charging, respectively.

### 2.1 Aerosol-cloud interactions

Aerosol fields in the model are prescribed using two distinct configurations. The first is an idealized setup, in which both CCN and IN are assumed to have a single chemical composition (e.g., ammonium sulfate for CCN) and spatially uniform distributions (Cheng et al., 2007, 2010). The second is a quasi-realistic configuration, in which aerosols consist of multiple components—such as organic carbon (OC), black carbon (BC), sulfate, sea salt, and dust—and are assigned spatially varying distributions based on observational data (Wang et al., 2017; Chen et al., 2018; Wang et al., 2024). In this configuration, hygroscopic components (sea salt and sulfate) act as CCN, dust particles serve as IN, and all species affect the radiation transfer.

When serving as CCN, aerosols are assumed to follow a trimodal lognormal size distribution (Cheng et al., 2007). Several predefined aerosol size spectra are provided, based on the classifications of Whitby (1978) and Jaenicke (1993). Users can adjust the spectral parameters (i.e., the number concentration, geometric mean radius, and geometric standard deviation of each mode) to implement customized or observation-based aerosol size distributions (Chen et al., 2015, 2016).

The cloud microphysics is parameterized with a two-moment scheme (Cheng et al., 2010), which predicts both the number and mass mixing ratios of five hydrometeors, including cloud droplets, rain droplets, cloud ice, snow, and graupel. The scheme explicitly calculates the activation of aerosol particles into cloud droplets using an embedded parcel model that advances in sub-timesteps rather than taking saturation adjustment, and tracks aerosol masses in different hydrometeors, enabling the simulation of aerosol removal through precipitation and recycling upon complete evaporation of cloud droplets (Chen et al., 2015). Meanwhile, the scheme considers the activation of giant CCN into raindrop embryos, thereby improving the representation of precipitation initiation in pristine marine environments. The homogeneous ice nucleation is parameterized following DeMott and Rogers (1990) and DeMott et al. (1994). The heterogeneous ice nucleation is calculated according to Meyers et al. (1992), where the ice nucleation rate depends on relative humidity while confined by the available cloud-ice number and IN concentration.

Previous studies have demonstrated that this treatment of aerosol-cloud interaction can effectively simulate both CCN and IN effects of aerosols. For example, Cheng et al. (2007) successfully reproduced the aerosol CCN effect and giant CCN effect in warm clouds; Wang et al. (2024) revealed that incorporating a relatively realistic IN spatial distribution (derived from MERRA-2 aerosol reanalysis) may improve the precipitation prediction of the 21•7 extreme precipitation events of Henan, China; Chen et al. (2018) and Song et al. (2019) demonstrated that both aerosol-cloud-radiation and aerosol-radiation interactions have significant impacts on the East Asia summer monsoon.

**2.2 Cloud-lightning interactions**

The lightning parameterization incorporates five prognostic electrical variables, i.e., the charge densities associated with five hydrometeor classes. These variables are dynamically coupled in real time with the cloud microphysical processes. For each hydrometeor species, the change of charge density is governed by four processes: advection, charge sources/sinks, charge transfer during class transitions, and sedimentation (Xu et al., 2014; Zhao et al., 2015). In each timestep, charge generation, charge transfer, and sedimentation associated with cloud microphysics are firstly calculated to update the charge distribution, which is then used to calculate lightning discharge; and finally, the charge density of each hydrometeor is advected along with the related hydrometeor.

The parameterization of charge generation primarily considers the non-inductive and inductive charging schemes from Gardiner et al. (1985) and Mansell et al. (2005). Extensive observational and laboratory studies have established that the non-inductive charging mechanism is due to the graupel-ice collisions between approximately -10°C and -20°C (e.g., Takahashi, 1978; Saunders et al., 1991). In the two-moment cloud microphysical scheme above, these collision processes are considered to generally result in rebound and thus do not participate in microphysical computations. Therefore, we incorporate the parameterizations of graupel-ice and graupel-snow collisions proposed by Lin et al. (1983) - specifically for calculations in non-inductive charging processes (see details in Appendix A). While ice-snow collisions may be relevant in the stratiform regions of mesoscale convective systems (Mansell et al., 2005), existing laboratory studies (e.g., Brooks et al., 1997) emphasize that non-inductive charging is highly sensitive to the riming rate and collision velocities. Given that snow particles typically exhibit lower densities, slower fall speeds, and lower rime accretion rates compared to graupel, their contribution to total charge separation is expected to be secondary. Furthermore, the lack of a robust empirical parameterization for this specific process justifies its exclusion from our current modeling framework.

For the considered graupel-snow and graupel-ice collisions, the charge generation per collision for non-inductive processes is constrained based on Ziegler (1991), with upper limits of 50 fC for graupel-snow collisions and 20 fC for graupel-ice collisions.

The discharge parameterization follows the approach of Marshall et al. (1995), with computations performed independently for each column in the model domain (Xu et al., 2014). At every model time step, the procedure begins by calculating the vertical electric field intensity within a column. This is achieved by solving the Poisson equation using the successive over-relaxation (SOR) iterative method, with the total charge vertical profile serving as input. The resulting electric field intensity is then evaluated at all vertical levels. If the field intensity at any level exceeds both the height-dependent theoretical breakdown threshold (Marshall et al., 1995) and a fixed minimum of 30 kV/m (Xu et al., 2014), a discharge is triggered. This discharge reduces the charge by 20% at all levels where the threshold is exceeded and is recorded as a single lightning event for that column. The entire procedure is iterated within the same time step until the electric field throughout the column falls below the breakdown threshold, thereby allowing for multiple discharges per time step. The scheme does not distinguish between cloud-to-ground and intra-cloud lightning.

## 3 Experiment Setup

This study investigates aerosol-cloud-lightning interactions through two thunderstorm cases. The first is an isolated severe thunderstorm in Guangdong province, which is used to validate the model capability in reproducing observed electrical structures. The second case is a frontal thunderstorm that occurred in June 2014 in Guangdong, aimed at quantifying the sensitivity of lightning activity to the perturbation of aerosol concentrations.

### 3.1 The isolated thunderstorm event

The simulation of the isolated thunderstorm was run for 24 hours from 1200 UTC on 15 June to 1200 UTC on 16 June 2021. The first 12-hour period was considered the model spin-up, leaving the subsequent 12-hour period for analysis. The analysis period captured the complete lifecycle of the isolated thunderstorm event, including its initiation and dissipation, which took place between 0800 and 1000 UTC on 16 June.

A two-way interactive nested-domain was employed, as illustrated in Figure 2. The parent domain (D01) spanned 301 × 400 grid points at 9 km horizontal resolution, while the nested domain (D02) encompassed 400 × 361 grid points at 3 km resolution to better resolve convective-scale processes. The model comprised 43 eta levels with the top at 50 hPa. Initial and lateral boundary conditions were derived from the 1° × 1° NCEP Final (FNL) Operational Global Analysis dataset (NCEP, 2000). The model time step was 30 seconds.

In the aerosol-cloud-lightning modeling framework, aerosols were prescribed using the idealized setup. The CCN concentration was assumed to be horizontally uniform, vertically exponentially decreasing with height. Over land and ocean grids, the aerosol size distributions were assumed to follow the marine and continental size spectra of Whitby (1978), respectively. The IN number concentration was set to a spatially uniform value of 150,000 #/m³. The aerosol radiative effects were disabled.

Other physical parameterizations were configured following Wang et al. (2024), including the Kain-Fritsch cumulus scheme (Kain, 2004; exclusively for Domain 1), the unified Noah surface scheme (Chen & Dudhia, 2001), the revised MM5 Monin-Obukhov surface-layer scheme (Jiménez et al., 2012), the YSU planetary boundary layer scheme (Hong et al., 2006), and the RRTMG longwave and shortwave schemes (Iacono et al., 2008).

To ensure the large-scale circulation better matches the observations, the spectral nudging (Storch et al., 2000; Choi & Lee, 2006) was applied to the outer domain (i.e., D01) during the spin-up phase. During the analysis period, model outputs were saved at 6-minute intervals to resolve high-frequency storm features.

The observational data of the charge structure are obtained from the study by Huang et al. (2025), where the spatial distribution of lightning radiation sources and their polarity identification were observed using the Guangdong Lightning Mapping Array (Huang et al., 2025). Note that these data only reflect charge regions that participated in the discharge process and inherently excluded those not involved in lightning discharges.

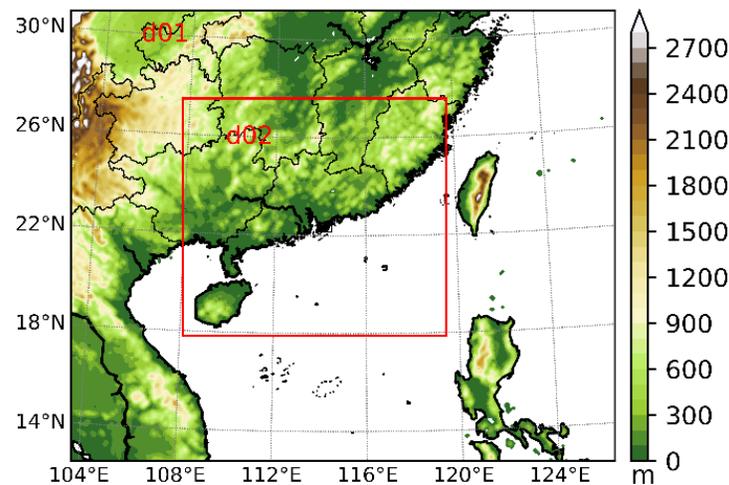

**Fig. 2.** Domain setup in WRF simulations. The shadings denote the topography (unit: m).

## 3.2 The frontal thunderstorm event

The frontal thunderstorm simulation employed the identical model configuration to that used for the isolated thunderstorm case. The frontal thunderstorm events spanned from 0000 UTC 2 June to 1200 UTC 3 June 2014, with model outputs recorded every 20 minutes. The first 12 hours were also discarded as the spin-up period, while the subsequent 24-hour period was used to analyze convective evolution and lightning activity.

To investigate the lightning response to varying aerosol loading, we conducted a set of sensitivity experiments: a control simulation (CTL) with aerosol conditions identical to that in the isolated thunderstorm event, a clean scenario with CCN concentrations reduced to 10% of the CTL level (CTLx0.1), and a polluted scenario with CCN concentrations increased to 10 times the CTL level (CTLx10).

For this event, the simulated surface precipitation was evaluated using the GPM (Global Precipitation Measurement) Level 3 IMERG Final half-hourly product (Huffman et al., 2023). This dataset features a spatial resolution of 0.1°×0.1° and utilizes an advanced algorithm that synergistically integrates observations from passive microwave sensors, infrared radiometers, and ground-based rain gauge networks.

The simulated lightning was evaluated using the lightning observations retrieved from GDLLS (Guangdong Lightning Location System), which detects cloud-to-ground (CG) flashes with high detection reliability (Chen et al., 2012). The system achieves a detection efficiency exceeding 90% and a spatial accuracy within 1 km across Guangdong Province (Zheng et al., 2016).

## 4 Validation of charge structure

Figure 3 compares the simulated temporal evolution of the charge vertical profile. As illustrated in Fig. 3a, the observed thunderstorm exhibited a tri-polar charge structure, characterized by a well-defined positive charge region in the upper layer (temperature > -20 °C), a primary negative charge region in the middle levels (around -10 °C), and a weaker positive charge region in the lower part. This structure (Fig. 3a) evolved through three stages. During the initial stage (08:06–08:18 UTC), the charge structure was relatively undeveloped. As the storm intensified (08:18–09:24 UTC), both the upper positive and mid-level negative charge regions were maintained at higher altitudes, with the lower positive charge region becoming actively involved in discharges. In the decaying stage (09:24–09:54 UTC), the altitudes of all charge regions decreased significantly, accompanied by the dissipation of the positive charge in low levels.

The simulation results demonstrated general high consistency with the observed charge vertical structure and temporal evolution, indicating the model capability of representing thunderstorm electrification processes across all three stages. Specifically, during the intensifying stage (08:18–09:24 UTC), the heights of the simulated charge centers aligned closely with the observed spatial distribution of lightning radiation sources. Likewise, the spatial extent and temporal evolution of the lower positive charge region during this intensifying stage were also reproduced with minimal deviation.

The main discrepancy between the simulation and the observation is the vertical displacement of the negative charge center. While the observed charge density was predominantly located within the -10 to -20°C temperature layer, the simulation located it in a relatively lower (warmer) level, mainly between the -10 and 0°C layer. This divergence is primarily due to the fundamental difference between the physical variables being compared. Owing to the limitations of current technology in directly measuring a continuous three-dimensional charge field, the observation must rely on lightning radiation sources, which are discrete points representing the initiation of electrical breakdown (Huang et al., 2025). In contrast, the simulation results are presented as horizontally averaged vertical profiles of charge density, characterizing the vertical distribution of net charges carried by hydrometeors. According to MacGorman et al. (2005), these discrete discharge events tend to initiate in regions of intense potential gradients at the boundaries of charge layers rather than filling the entire volume uniformly. Therefore, a spatial offset is physically expected between the center of the continuous simulated charge field and the clusters of discrete observed radiation sources.

Overall, it is concluded that the modeling framework can well simulate the observed vertical structure of charge fields in the isolated thunderstorm event.

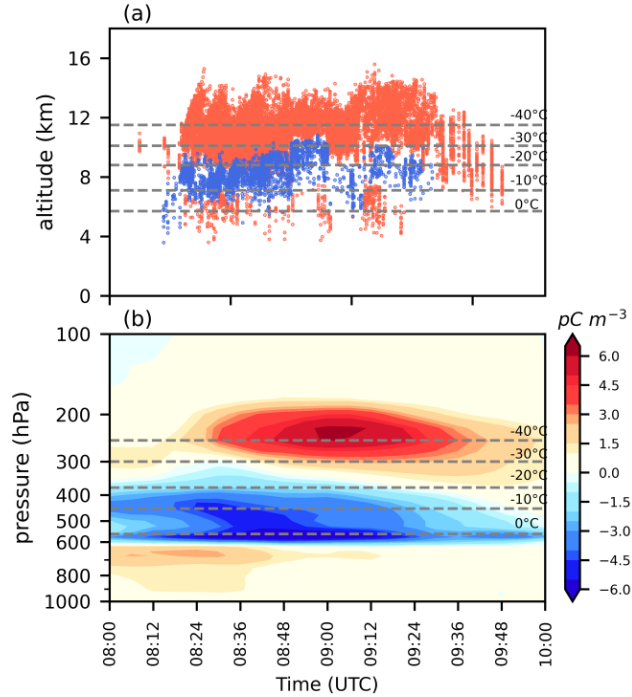

**Fig. 3.** (a) Time-height distribution of region-averaged (113°E–114°E and 23°N–24°N) of lightning radiation sources with different polarities in the observation. Orange (Blue) color denotes negative (positive) polarity breakdown in positive (negative) charge regions. The lightning radiation source detection and localization data were sourced from Huang et al. (2025). (b) Corresponding distribution of total charge density (pC m⁻³) in the simulation.

## 5 Aerosol effects on precipitation and lightning

This section first analyzes observed and simulated spatiotemporal patterns of precipitation and lightning in the frontal thunderstorm event, and examines their responses to aerosol concentration variations using sensitivity simulations. Subsequently, we investigate the microphysical mechanisms by which aerosols exert their influence on lightning activity.

### 5.1 Comparison of observed and simulated precipitation and lightning

Figure 4 compares observed (OBS; Fig. 4a,4c) and simulated (CTL; Fig. 4b, 4d) spatial distributions of the surface precipitation and lightning intensity for the frontal event, averaged over the analysis period (i.e., 24 hours between 1200 UTC 2 June and 1200 UTC 3 June 2014). Observational results show that the precipitation is primarily concentrated near 114°E (Fig. 4a). The domain is dominated by widespread light rainfall (<50 mm day$^{-1}$), interspersed with scattered spots of heavy precipitation (>100 mm day$^{-1}$).

The comparison reveals that the model reproduces the large-scale spatial pattern of the observed precipitation, though certain spatial displacements of intense convective centers are evident (Fig. 4b vs. Fig. 4a). The spatial correlation coefficient (PCC) is 0.36, which reflects a systematic spatial displacement and a more fragmented distribution of simulated convective cells compared to the observations (Fig. 4). Nevertheless, the model's skill in predicting precipitation above a 10 mm threshold, as measured by the Threat Score (TS10), is 0.63. This relatively high TS10, combined with the successful replication of the northeast-southwest gradient, demonstrates that the simulation possesses a reliable capability to characterize the morphological structure and intensity of the storm system, providing a physical basis for investigating aerosol-lightning interactions.

Regarding the lightning activity, the observations reveal a high-density lightning region (>1.50 km$^{-2}$ h$^{-1}$) within the domain. The lightning density exhibits a northeast-southwest gradient. The simulation effectively reproduces the spatial distribution of lightning activity, particularly in high-density lightning regions, although it underestimates both the frequency and intensity of lightning in low-density regions (< 0.06 km$^{-2}$ h$^{-1}$).

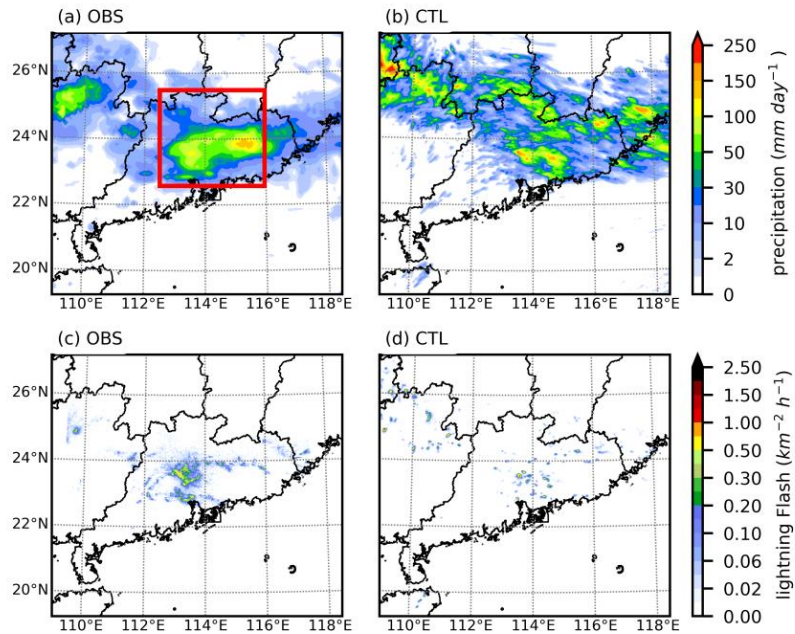

**Fig. 4.** Spatial distribution of 24-hour accumulated precipitation rate (a–b) and averaged CTL lightning density (c–d) for the period of 12 UTC 02 June to 12 UTC 03 June 2014 from the observation (left) and the CTL simulation (right). The red solid rectangle delineates the heavy-rainfall region (112.5°E–116°E and 23°N–25.5°N).

Figure 5 compares the temporal variations of the regionally-averaged precipitation rate and lightning flash density over the heavy-precipitation region (112.5°E–116°E and 23°N–25.5°N; the red rectangle in Fig. 4a) from 12 UTC 2 June to 12 UTC 3 June 2014 between observational data and the CTL simulation.

The simulated precipitation (Fig. 5a) shows reasonable agreement with observations, capturing the peak timing and the temporal evolution. While the simulation slightly overestimates the initial precipitation rate, it provides a consistent representation of the heavy rainfall event's evolution.

For lightning flash density (Fig 5b), the observation shows that the lightning activity peaked at 22:00 UTC, which is approximately one hour before the peak of the precipitation rate (23:00 UTC). This time gap—where lightning peaks before the maximum rainfall—aligns with the physical evolution of deep convection. It reveals that intense electrification driven by strong updrafts typically occurs before the maximum hydrometeor fallout reaches the surface (Williams et al., 1989). Nevertheless, the CTL simulation captures the later secondary peaks (e.g., after 04:00 UTC).

Notably, the CTL simulation captures the essential temporal evolution of lightning density, despite an underestimation in absolute magnitude. It is important to note that current discharge parameterization schemes are primarily valued for their capacity to capture trends rather than estimate precise lightning flash counts. This perspective is supported by the work of Marshall et al. (1995), which demonstrates that while lightning initiation follows physical thresholds (i.e., the breakeven field) related to electric field magnitude, the inability of models to capture sub-grid scale electric field variability makes absolute quantification challenging. Furthermore, the simulated lightning frequency is inherently sensitive to the parameter selection within electrification and discharge schemes (Mansell et al., 2005).

The results indicate that the lightning density in the CTLx10 simulation is approximately 1.5 times that of the CTLx5 simulation, representing a 50% enhancement in lightning activity as the CCN concentration doubles. This magnitude of response is in good agreement with the findings of Sun et al. (2021). Specifically, the flash density in the CTLx5 simulation shows a much closer alignment with the observed intensity (Fig. 5b). Given that the model calculates total lightning (incorporating both intra-cloud and cloud-to-ground flashes) while the observations only record cloud-to-ground (CG) flashes, the higher magnitude in CTLx5 is physically consistent with the documented total-to-CG ratio of approximately 3:1 or 4:1 (Prentice & Mackerras, 1977). This underscores that aerosol number concentration is essential for simulating the lightning density, suggesting that the initial bias in the CTL simulation may stem from the initial prescribed aerosol concentrations.

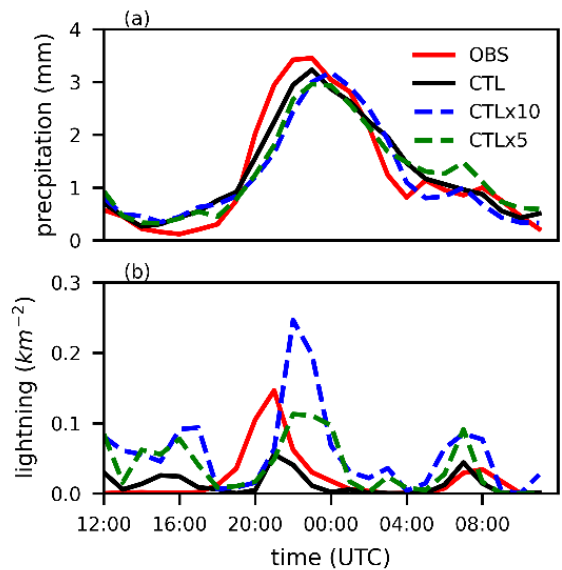

**Fig. 5.** Time series comparison of observed and CTL-simulated (a) precipitation and (b) lightning flash density from 12 UTC 02 June to 12 UTC 03 June 2014. The lightning and precipitation were both calculated over the region shown in Figure 4a, the area between 112.5°E–116°E and 23°N–25.5°N.

Figure 6 shows the spatial distributions of precipitation and lightning density from the CTLx0.1 and CTLx10 sensitivity simulations. The results indicate that increasing CCN concentrations suppress the surface precipitation (Figs. 6a, 6b vs. Fig. 4b), characterized by reduced precipitation intensity and less heavy-rainfall cores. Conversely, increasing CCN concentrations results in more intense lightning activity and more extensive spatial coverage of lightning density (Figs. 6c, 6d vs. Fig. 4d).

These are consistent with the results of previous studies and further confirm that higher CCN concentration tends to suppress precipitation while enhancing lightning activity. These results are consistent with several previous studies and suggest that higher CCN concentrations (e.g., exceeding 2,000 cm$^{-3}$) often tend to suppress surface precipitation, although this effect can be modulated by environmental factors such as relative humidity (Cheng et al., 2010; Barrett et al., 2023). This suppression is conventionally attributed to the formation of more numerous but smaller droplets, which inhibits warm-rain formation by reducing collision-coalescence efficiency. However, in mixed-phase clouds, the response is more complex: while some studies suggest an enhancement of ice-phase processes, Barrett et al. (2023) demonstrated that a significant increase in CCN can actually lead to a drastic reduction in riming collection efficiency (e.g., from 0.79 to 0.24). Furthermore, as highlighted by Cheng et al. (2010), under low RH conditions, the increased surface area of smaller droplets facilitates rapid evaporation, further intensifying precipitation suppression. In contrast, the mechanisms underlying the enhancement of lightning activity are less understood. The following section will therefore focus on how aerosols modulate lightning activity through microphysical pathways.

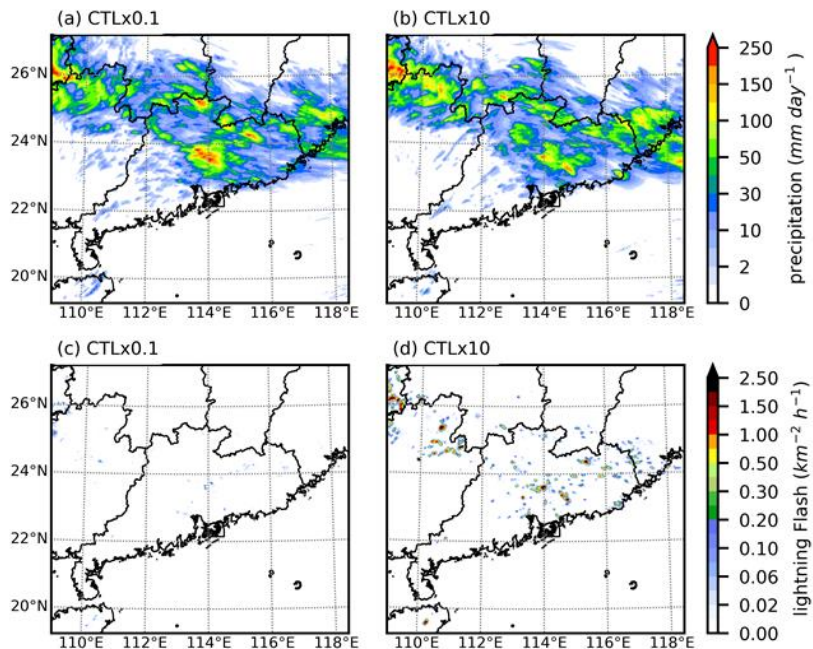

**Fig. 6.** As Figure 4, but for results of CTLx0.1 (a, c) and CTLx10 (b, d) simulations.

### 5.2 Underlying mechanism of aerosol effects on lightning activity

Figure 7 compares the simulated temporal variations of regionally-averaged vertical profiles of cloud ice, snow, and graupel mass mixing ratios across the three WRF simulations. These ice-phase hydrometeors are crucial to charge generation in clouds. The simulations reveal that higher CCN concentrations lead to larger cloud water contents (contours in Figs. 7a-7c), which in turn induce marked changes in ice-phase cloud microphysics. Specifically, the cloud ice mixing ratio increases as more cloud droplets are lofted into upper layers. In contrast, the graupel mixing ratio decreases evidently, suggesting suppressed riming efficiency. The snow mixing ratio exhibits a more complex response: it decreases during the early analysis period (i.e., around 12:00 UTC) but increases later, particularly between 20:00 and 04:00 UTC. On average, over the whole analysis period, the change in snow mix ratio is minor.

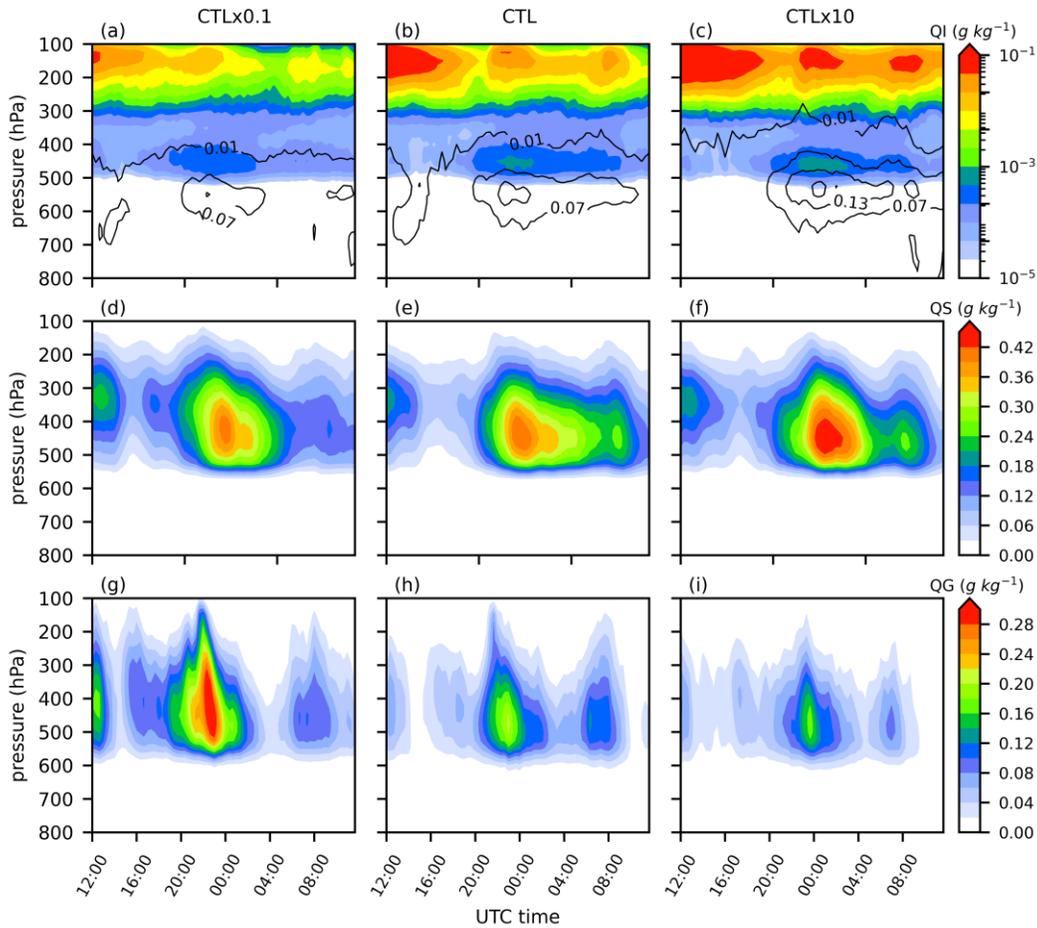

**Fig. 7.** Temporal-vertical variations of cloud ice (a–c), snow (d–f), and graupel (g–i) mass mixing ratios in CTLx0.1 (left column), CTL (middle column), and CTLx10 (right column) simulations, averaged over the red rectangle shown Fig. 4a. Black lines in the upper row denote contours of simulated cloud water mixing ratios (units: g kg$^{-1}$).

Accompanied by changes in ice-phase cloud microphysics, the charge densities of various hydrometeors are altered correspondingly. As shown in Fig. 8a-c, the region of positive changes associated with cloud ice in the upper atmosphere is significantly expanded as the CCN concentration increases. It should be noted that this enhancement of upper-level positive charge primarily shows the vertical transport of cloud ice particles, rather than charge generation at those altitudes. The charge generation in our simulations is dominated by the non-inductive charging mechanism (Takahashi, 1978; Saunders et al., 1991) associated with graupel-ice collisions (Eq. A2), which occurs predominantly in the mid-levels at approximately 450–350 hPa (see Appendix A). The graupel-ice charge separation is enhanced by two key factors: an increased cloud ice mass mixing ratio, which raises the graupel-ice collision rate ($P_{GACI}$ in Eq. A2), and a reduction in cloud ice size ($XM_{ice}$ in Eq. A3), which improves separation efficiency. Although a reduction in graupel size acts to increase the size parameter ($\lambda_G$ in Eq. A2) and suppress $Q_{GACI}$, this effect is minor compared to the dominant influence of cloud ice changes, given the limited range of mean graupel radius variation (CTLx0.1: 0.456 mm, CTL: 0.423 mm, CTLx10: 0.430 mm).

The resulting enhancement in charge separation produces a greater abundance of positively charged ice particles. These particles are subsequently transported by updrafts to the ice-rich regions of the upper atmosphere, leading to the expansion of the positive charge zones observed in Figs. 8a–c. The intensified charge separation leads to a higher positive charge density on cloud ice. These cloud ice particles are subsequently carried upward by updrafts to the upper levels (100–300 hPa), resulting in a prominent expansion of the upper-level positive charge zones shown in Figs. 8a–c. In contrast, the charge densities associated with snow and graupel remain relatively unchanged (Figs. 8d-i). This is consistent with their primary distribution in the lower and middle layers, where they are less susceptible to vertical transport.

Figure 8 represents the charge structure after discharges, as the model outputs the net charge density after the neutralization process in each lightning-triggering time step. Although the CTLx10 simulation experiences more vigorous electrification, its significantly higher lightning frequency leads to more intense neutralization. As evidenced by the average vertical profiles in Fig. 9d, the potential gradient exhibits a consistent increase from CTLx0.1 to CTLx10, confirming that higher aerosol concentrations strengthen the cloud's electrification, maintaining a higher electrical gradient.

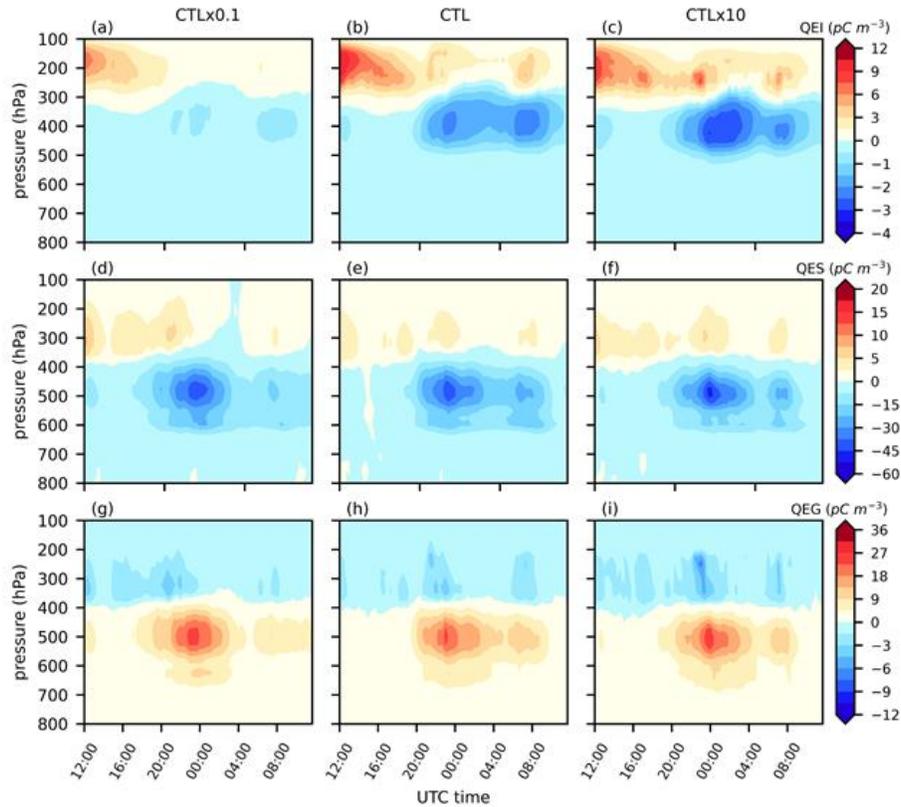

**Fig. 8.** Same as Fig. 7, but for charge densities (units: pC m$^{-3}$) associated with cloud ice (QEI), snow (QES), and graupel (QEG), respectively.

Figure 9 compares the vertical profiles of total charges in three simulations, averaged throughout the analysis period. Clearly, the tripolar charge vertical structure is formed by the interplay of three ice-phase hydrometeors. The upper positive region is dominated by cloud ice, the middle negative region is governed by snow, and the lower weakly-positive region corresponds to graupel. As the CCN concentrations increase from 0.1 to 10 times the background levels, the upper positive-charge region intensifies markedly, with peak values rising from 6 (Fig. 9a) to 12 pC m$^{-3}$ (Fig. 9c), and expands spatially. Shown in Fig. 9d, increasing aerosol concentrations simultaneously intensify both the positive charge region at 100-300 hPa and the negative charge region at 400-600 hPa, leading to an enlarged electric potential gradient between these layers. This enhanced gradient is most significant within the lightning hotspot region centered at (114°E, 23.5°N).

Several observations and numerical studies have indicated that an increase in CCN concentration promotes the growth of ice-phase particle mass mixing ratios, thereby enhancing lightning activity (Rosenfeld et al., 2008; Wang et al., 2011; Fan et al., 2013). However, Tan (2017) pointed out that as aerosol concentrations increase, the mixing ratio of graupel particles tends to decrease. Furthermore, Sun et al. (2021) observed that even with a reduction in graupel concentration, lightning activity is still enhanced with CCN concentration increasing, primarily due to an increase in cloud ice numbers and an enlargement of graupel particle radius. The changes in graupel particles observed in this study align well with the findings of Sun et al. (2021).

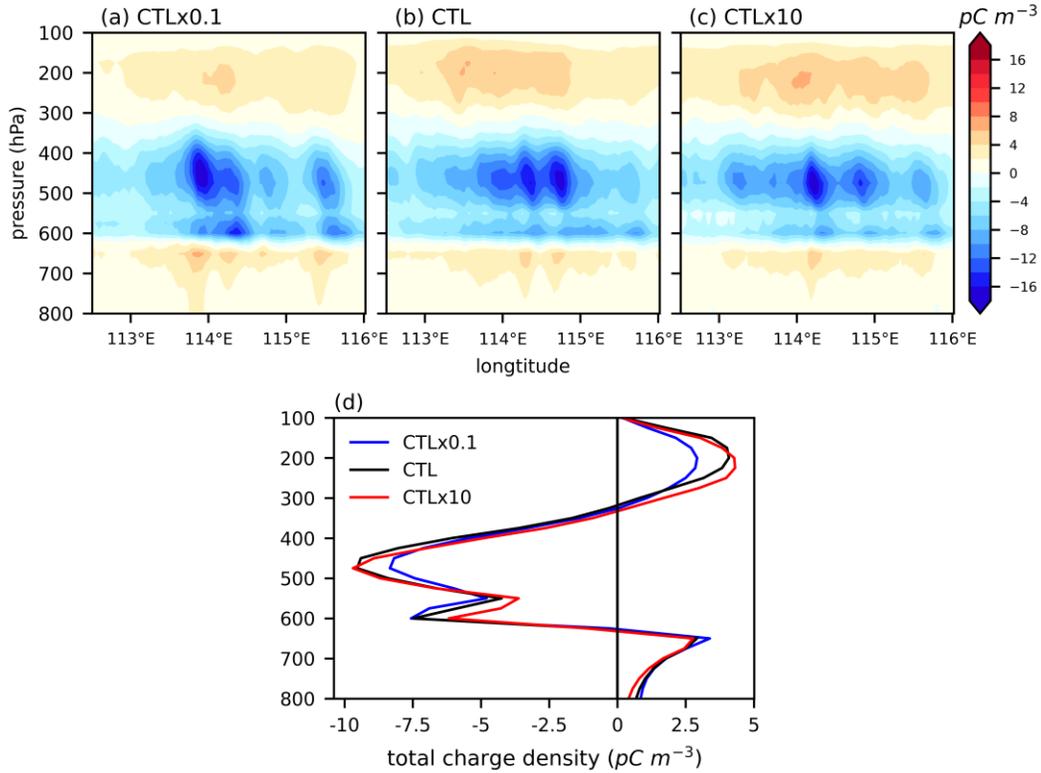

**Fig. 9.** Zonal variation of meridionally-averaged, 24-hour mean charge structure under different CCN concentrations: (a) CTLx0.1, (b) CTL, and (c) CTLx10. Shading indicates the magnitude of charge density (pC m$^{-3}$). (d) Corresponding vertical profiles of the total charge density, averaged over the red rectangle denoted in Fig. 4.

Figure 10 summarizes the identified mechanism using two radar charts. As CCN concentration increases, the cloud droplet number concentration rises while the droplet size decreases, suppressing warm-rain processes, increasing cloud water content, and enabling more cloud water to be lofted into the upper troposphere (Fig. 10a). This, in turn, enhances homogeneous ice nucleation, resulting in a larger number of smaller cloud ice particles and greater cloud ice mass. These changes subsequently lead to a decrease in graupel mass and a slight increase in snow mass.

These microphysical alterations, particularly the increase in cloud ice mass, trigger a redistribution of charge among hydrometeors (Fig. 10b). The increase in cloud ice mass provides an abundance of small cloud ice that can be transported into regions containing riming graupel, intensifying graupel-ice collisions and the associated non-inductive charge separation. This process results in cloud ice acquiring a stronger positive charge, while the reduction in the negative charge on snow is relatively small. Consequently, an enhanced vertical electric field gradient develops between the upper and middle cloud layers, which in turn promotes lightning activity.

We also examined convective intensity across the three simulations, as previous studies have suggested that aerosols may enhance lightning activity by invigorating convection. However, the mean updraft velocity (w > 0 m/s) shows a minor difference across the CTLx0.1, CTL, and CTLx10 cases (0.189, 0.176, and 0.171 m s$^{-1}$), while the 99th percentile maximum updraft velocity reaches 2.192, 1.956, and 1.844 m s$^{-1}$. Therefore, we conclude that aerosol-induced changes in cloud microphysics, rather than convective invigoration, dominate the modulation of lightning activity.

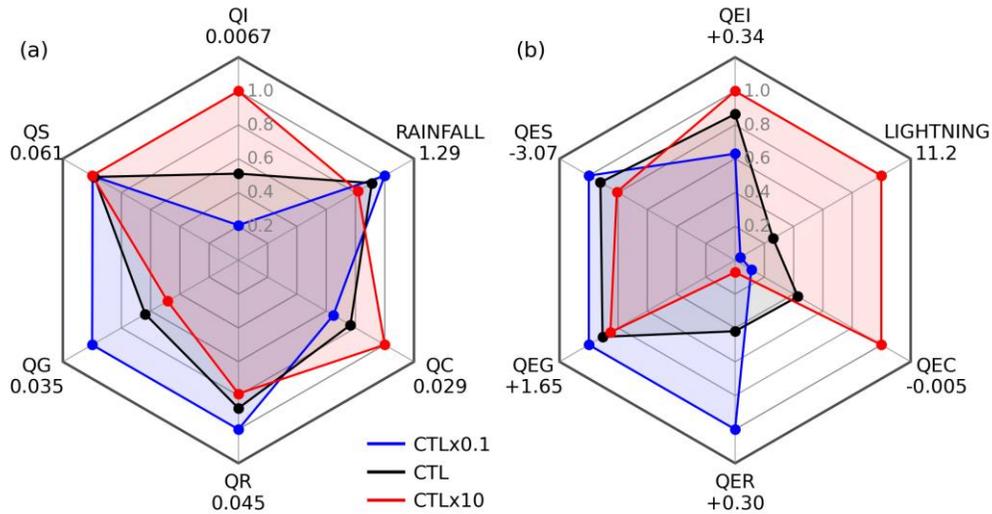

**Fig. 10.** Radar chart of the hydrometeors mixing ratios (unit: g kg$^{-1}$) and surface rainfall rate (unit: mm h$^{-1}$) (a) and that of the vertically-averaged charge densities (unit: pC m$^{-3}$) and lightning density (unit: km$^{-2}$) (b) under different CCN conditions. Results are spatial-temporally averaged over the 24 hours within the region indicated in Fig. 4a. All axes are normalized to the maximum value among the three cases, with the absolute maximum values labeled below each axis name. CTLx0.1, CTL, and CTLx10 represent experiments with 0.1, 1, and 10 times the control CCN concentration, respectively.

**6 Summary and Discussion**

This study develops a novel framework for modeling aerosol-cloud-lightning interactions within the WRF model. The framework facilitates investigating the effects of multi-component aerosols on cloud microphysical processes and their subsequent impact on lightning activity. It is validated against observations that the framework successfully reproduces the characteristic tripolar charge structure of an isolated thunderstorm. Since cloud microphysical processes are often difficult to observe directly, their parameterizations remain challenging to verify. In contrast, lightning observations provide a surrogate means of constraining microphysical processes. Therefore, this framework, combined with lightning observations, offers an effective approach for validating and improving the representation of ice-phase microphysics in models.

The framework also simulates that increasing aerosol concentrations tends to intensify the lightning activity, consistent with previous studies. However, unlike most prior work—which attributes lightning enhancement primarily to the aerosol-induced convective invigoration—this study suggests that the increase in lightning is more closely linked to changes in cloud microphysics. Specifically, higher aerosol concentrations enhance cloud water content and lead to greater cloud ice mass, which in turn intensifies graupel–ice collisions and the associated non-inductive charge separation. This strengthens the vertical electric field gradient, thereby yielding increased lightning activity. While these mechanistic findings are valuable, they are derived from two thunderstorm cases in Guangdong, China. Given that the lightning response to aerosols can vary significantly across different meteorological regimes, the generalizability of these results to continental, maritime, or mountainous environments requires further validation. Expanding the framework to include broader case studies in future work would strengthen the robustness of these conclusions.

While the framework advances the modeling of aerosol-cloud-lightning interactions, it can still be improved in several aspects. First, the model incorporates only the non-inductive charging scheme proposed by Gardiner et al. (1985) and Ziegler et al. (1991). The scheme has difficulties in replicating observed inverted-polarity charge structures, which may suppress positive charging efficiency within the crucial -10°C to -20°C zone (Mansell et al., 2005). Incorporating additional and perhaps more sophisticated non-inductive schemes, such as the riming rate-dependent scheme (Saunders & Peck, 1998) or the effective water content-dependent scheme (Saunders et al., 1991), would enable the simulation of a wider range of electrification scenarios. Second, a more detailed discharge parameterization is needed to better simulate the spatial and temporal evolution of lightning flashes. A promising solution is to merge two complementary approaches: the physically-constrained discharge model from Fierro et al. (2013) for determining lightning initiation and type, and the explicit three-dimensional stochastic breakdown model from Mansell et al. (2002) for simulating the bidirectional propagation and branching of lightning leaders. Ultimately, these refinements will not only enhance the model predictive skill but also deepen our understanding of the complex interplay between aerosols, cloud dynamics, and electrical activity in the atmosphere.

It is worth noting that while the current model framework is capable of incorporating both aerosol-radiation interactions (ARI) and ice nucleating particle (INP) effects, these processes were not activated in the present simulations to isolate and emphasize the CCN-mediated microphysical pathways. Given the complexity of aerosol-cloud-lightning interactions, we aim to establish a baseline understanding by focusing on CCN-mediated effects. These mechanistic findings carry profound implications for weather forecasting amidst large-scale emission reductions. In transitioning environments like China, a reduction in aerosol concentrations diminishes the cooling effect of ARI, thereby enhancing atmospheric instability. Under such conditions, even as total aerosol loading decreases, concentrations often remain sufficient to suppress the warm-rain process, allowing CCN-mediated microphysical pathways to exert influence on the intensification of lightning activity. The influences of INP and aerosol-radiation interactions will be addressed as a focal point of our follow studies, leveraging the capabilities of the current model framework.

**Acknowledgements**

This study is supported by Guidance Project for Industrial Technology Development and Application Plan of Fujian Province (2024Y0076) and National Science Foundation of China (U2342215).

**Appendix A: Parameterization of graupel collection rates for snow and cloud ice**

The collection rates of graupel particles collecting cloud ice ($P_{GACI}$) and snow ($P_{GACS}$) are expressed as (Lin et al., 1983):

$$P_{GACS} = \pi^2 E_{GS} n_{0S} n_{0G} |U_G - U_S| \frac{\rho_s}{\rho} \times \left[ \frac{5}{\lambda_S^6 \lambda_G} + \frac{2}{\lambda_S^5 \lambda_G^2} + \frac{0.5}{\lambda_S^4 \lambda_G^3} \right] \quad (A1)$$

$$P_{GACI} = 0.25\pi \frac{n_{0G} E_{GI} Q_I * \Gamma(3.5)}{\lambda_G^{3.5}} \times \left( \frac{4g\rho_G}{3C_D \rho} \right)^{0.5} \quad (A2)$$

In these equations, $\rho$, $\rho_s$ and $\rho_G$ represent the densities of air, snow, and graupel, respectively. The collection efficiencies ($E_{GS}$, $E_{GI}$) and the intercept parameters ($n_{0S}$, $n_{0G}$) for graupel and snow are prescribed as constants. $U_G$ and $U_S$ are their terminal velocities. $Q_I$ is the cloud ice mass mixing ratio. An increase in the slope parameters ($\lambda_G$, $\lambda_S$) leads to a reduction in the mean particle size for graupel and snow, respectively. The drag coefficient of graupel particles ($C_D$) is a constant. $\Gamma$ denotes the gamma function with $\Gamma(3.5) \approx 3.323$, and g is gravitational acceleration (9.8m s$^{-2}$).

The term $\frac{\partial \rho_{GI}}{\partial t}$ represents the non-inductive charge rate between cloud ice and graupel (Mansell et al., 2005), is defined by the following equations:

$$\frac{\partial \rho_{GI}}{\partial t} = \beta \delta Q (1 - E_{GI}) \times E_{GI}^{-1} \times \left( \frac{P_{GACI}}{XM_{ice}} \right) \quad (A3)$$

The charge transferred per collision ($\delta Q$) is parameterized by the GZ scheme (Gardiner et al., 1985; Ziegler et al., 1991), which is empirically derived from laboratory measurements

(Jayaratne et al.,1983). Key parameters include the charge separation limiting coefficient β (0 ⩽ β ⩽ 1), the mean cloud ice particle mass ($XM_{ice}$). T represents temperature.

$$\delta Q = k D_k^4 (\delta v)^3 \, \delta L f(\tau) \qquad (A4)$$

$D_k$ denotes the diameter of cloud ice or snow particle, while δv represents the terminal velocity difference between colliding particles. δL is a function of liquid water content, and f(τ) incorporates both reversal and ambient temperature (Ziegler et al., 1991).

**Open Research**

Figures were made with Matplotlib version 3.2.1 (Caswell et al., 2020; Hunter, 2007), available at https://matplotlib.org/.